\documentclass{iau} 
\usepackage{graphicx}

\newcommand{\apj}{{\it ApJ}}
\newcommand{\solphys}{{\it Sol.~Phys.}}
\newcommand{\icarus}{{\it Icarus}}
\newcommand{\apjl}{{\it ApJ Letters}}

\title[JD 11.~~Faint Young Sun Paradox] 
{The Faint Young Sun and\\ Faint Young Stars Paradox}

\author[Petrus C. Martens]   
{Petrus C. Martens}

\affiliation{Department of Physics \& Astronomy,\\
Georgia State University,\\
25 Park Place, 6$^{th}$ Floor,\\
Atlanta, GA 30303, USA\\ 
email: {\tt martens@astro.gsu.edu}}

\pubyear{2017}
\volume{328}  
\jname{Living Around Active Stars}
\editors{D. Nandi, A. Vallo, P. Petit, eds.}
\begin{document}

\maketitle

\begin{abstract}

The purpose of this paper is to explore a resolution for the Faint Young Sun Paradox
that has been mostly rejected by the community, namely the possibility of a somewhat
more massive young Sun with a large mass loss rate sustained for two to three billion
years.  This would make the young Sun bright enough to keep both the terrestrial and
Martian oceans from freezing, and thus resolve the paradox.  It is found that a large and
sustained mass loss is consistent with the well observed spin-down rate of Sun-like
stars, and indeed may be required for it.  It is concluded that a more massive young Sun
must be considered a plausible hypothesis.

\end{abstract}

\firstsection 
\section{Introduction}

The young Sun started its life on the main sequence with about 70\% of the
luminosity of what it has now according to standard stellar evolution theory.
It is still a scientific riddle how, with such a faint Sun, the young Earth could be
warm enough to host liquid water in its first couple of billion years.  Yet
geological evidence clearly indicates there have been warm oceans from very
early on (Kasting 1989), and that these oceans were a key ingredient in the
development of life.

This is called the Faint Young Sun Paradox.  The paradox is even more
compelling for the planet Mars which we know now to have been covered
with oceans for periods of hundreds of millions of years in its early life, with
only half of the incoming energy flux of sunlight of the Earth.

Stellar evolution simulations dictate his paradox and it therefore applies to all G
stars, and less so for K and M stars that evolve much slower. 
In all cases the habitable zone around
the star gradually moves outwards and planets that started out balmy are
expected to end up scorched.  Given that it took about four billion years on
planet Earth for the development from single cell organisms to multi-cellular life
-- and since that is the only example of evolution we have -- it is a reasonable
assumption that the development of multicellular intelligent life takes a very long
time in general, with most G star planets not spending enough time in the habitable
zone.

This paradox has been known for a long time, and one of the first to hint at 
a solution was well known science popularizer Carl Sagan (Sagan \& Mullan 1972). 
Many solutions
to the Faint Young Sun Paradox have been proposed over the years, and they
come from very different fields.  Fairly straightforward proposals are an enhanced
greenhouse effect by carbon dioxide or methane, geothermal heath
from an initially much warmer terrestrial core, a much smaller Earth albedo, life
developing in a cold environment under a 200 meter thick ice sheet, a secular
variation in the gravitational constant, etc.  Most of these models have serious
shortcomings:  For example the greenhouse effect from methane appears to be
self-limiting, and not enough CO2 is indicated by the geological record to justify
a greatly enhanced greenhouse effect in the past (Kasting, 2004)

There is not enough space in a proceedings paper to review all the material
discussed above, so I refer to a recent review by Feulner (2012)  and a series of
very enlightening presentations and papers by Dr.\ James Kasting (Kasting, Toon
\& Pollack 1988; Kasting, 2004) on the many hypotheses
proposed to resolve the Faint Young Sun Paradox.

Most of the solutions proposed for the Faint Young Sun Paradox apply to Earth
alone; they do not explain the presence of liquid oceans on early Mars.  Also, they
are not solutions for the Faint Young Stars Paradox in general.
A simpler solution has been proposed in
that the early Sun was more massive and hence more luminous.  This necessitates
a massive, sustained solar wind for the first billions of years of the Sun's evolution,
a condition that most in the community find implausible.

In the current paper I will explore the hypothesis of a more massive and hence
much brighter young Sun in more detail, and investigate whether this leads to logical
contradictions or can be ruled out by observations.  I will find, surprisingly, that 
a more massive young Sun is not implausible at all, and links together what we 
know about stellar spin-down with simulations of the same.
The results of this paper do not prove
that the young Sun was more massive -- we would need observations that
demonstrate the sustained presence of a massive solar wind.  But it does show
that such a hypothesis cannot be ruled out at present, and consequently that
the presence of oceans on early Mars may not be a conundrum, and that the
habitable zones around solar analogs as well may remain in place for the billions
of years it takes for multi-cellular and intelligent life to develop.

\section{Mass Loss and Luminosity}

A slightly more massive Sun would be significantly more luminous:  In the solar
portion of the Hertzsprung-Russel diagram luminosity scales with mass to the 
power 4 to 5.  If the Sun were more massive earlier on the Earth would be closer
in as well:  Because of conservation of it angular momentum the mean Sun-Earth
distance varies as the inverse of the mass of the Sun, while the incoming radiation
at the top of the Earth's atmosphere scales with the inverse of the square of that
distance.  Hence the amount of radiation the Earth receives varies as the mass of
Sun to the power 6 to 7.  A 30\% less luminous Sun at the Zero Age Main Sequence
(ZAMS) at one solar mass could be compensated for by a mere 4 to 5\% mass
increase going back in time from the current Sun.

A sustained solar mass loss rate of roughly 10$^{-11}$ M$_{\odot}$ yr$^{-1}$ is required
to accomplish that.  The current solar mass loss rate is estimated
at 2-3$\times$10$^{-14}$ M$_{\odot}$ yr$^{-1}$ for the fast wind and 
10$^{-15}$ M$_{\odot}$ yr$^{-1}$ for Coronal Mass Ejections (CMEs).  Interestingly
the mass loss from photon emission is twice as large, around 
7$\times$10$^{-14}$ M$_{\odot}$ yr$^{-1}$, but the latter contributes much less to
angular momentum loss, because the photons are not forced to co-rotate with the
magnetic field.

So the current mass loss rate, if extrapolated to the past, is insufficient to resolve
the Faint Young Sun Paradox by a factor of 300 or more.  Hence we must assume
a much higher mass loss rate for the young Sun, sustained for several billions of
years.  Observations of some young solar-type stars indicate mass loss rates of
roughly the right magnitude:  e.g.\ 70 Opiuchi with a mass of 0.92 times that of
the Sun, and an estimated age of 0.8 billion years, has a mass loss rate of 
3$\times$10$^{-12}$ M$_{\odot}$ yr$^{-1}$, and $\kappa^{-1}$ Cet
(Do Nascimento et al.\ 2016) yields the same result from X-ray calibration.

A much more detailed analysis than the back-of-the-envelope
calculation above, by Minton \& Malhotra (2007), narrows down the mass loss rate
constraint further.  Minton \& Malhotra calculate the solar mass and hence 
mass loss rate required to keep the radiative equilibrium temperature of the
Earth's atmosphere at 273$^{\circ}$ K, the freezing point of water, during solar
evolution.  It is plausibly assumed that the greenhouse effect will add about
15$^{\circ}$ K to that, as it does at present, to achieve the average atmospheric
temperature we have now, that is favorable to life.  The result of their analysis
is, again, a required mass loss rate of about 10$^{-11}$ M$_{\odot}$ yr$^{-1}$,
but it only has to be sustained for the first 2.4 billion years.

The choice of maintaining the radiative equilibrium temperature at or above
freezing is a rational one, because at a lower temperature planet Earth could
flip to an equilibrium in which all of the surface is frozen over -- snowball
Earth -- where the albedo is much higher, because of all the ice and snow.  The
geological record indicates that several
``snowball Earth" episodes have occurred in Earth's history -- in addition to the
much more recent ice ages, where there is no full planetary ice coverage.

As an aside, but in response to an obvious question:  How does the Earth's
atmosphere escape from a ``snowball Earth" state?   The answer probably
lies in the addition of CO2 to the atmosphere from volcanic eruptions.  A
slow but steady addition of CO2 by volcanism and no uptake of CO2 by the
weathering of rocks and diffusion into the oceans -- all covered by ice --
will eventually create
enough of a greenhouse effect to initiate melting at equatorial latitudes,
after which, via various feedbacks, melting will proceed precipitously. 

Minton \& Malhotra also point out that their model for ``minimum mass loss", 
according to the simulations of Kasting (1991), maintains solar luminosity at
a high enough level to keep the atmosphere of Mars above the freezing point
for the first billion years of its history -- when oceans are believed to have
existed on Mars.  So indeed a strong early solar wind can resolve the
paradox for both Earth and Mars, no separate solutions are required,
much to the liking of Father William of Occam.

\section{Stellar Spin Down and Mass Loss}

In this section I will relate stellar mass loss rates, which are hard to observe,
with the much better known stellar spin down rates in order to verify whether
these can be made consistent.

It is well known that Sun-like stars spin down from rotation periods of just a
few days in their first billion years to several weeks in their mid-life, e.g.\ 26
days for the Sun at 4.5 billion years.  The loss of angular momentum is
usually ascribed to the torque applied by the stellar wind that co-rotates
with the star near the surface and is forced to co-rotate roughly out to the
Alfv\'{en} radius where the wind outflow velocity equals the Alfv\'{en} speed.  

Weber \& Davis (1967) were the first to relate spin-down rates to a stellar
wind model.  Their model is of a purely radial magnetic field that changes
polarity at the equator.  Their key result 
that the torque applied by the wind on the star is to a good approximation
given by
\begin{equation}
T = \omega R_{A}^{2}  \dot{M},    
\end{equation}
where $\omega$ is the angular rotation rate of the star, $R_{A}$ the
critical Alfv\'{e}n radius where the wind outflow velocity equals the 
local Alfv\'{e}n speed, and $\dot{M}$ is the stellar mass loss rate.  The
calculation of Weber \& Davis includes at factor 2/3 on the right hand
side resulting from the azimuthal integration of the torque, which I omit
here for simplicity

Physically interpreted this means that the stellar wind torque is roughly
equal to the angular momentum of a  stellar wind forced to co-rotate up
to $R_{A}$ and then let go, flowing out further preserving its angular
momentum.  The result of
Eq.\ (3.1) follows from the requirement that the solution flows smoothly
through the critical point in the defining equations, much like the 
requirement for the critical point in the thermally driven Parker wind.

The location of the critical Alfv\'{e}n radius in the Weber \& Davis solution is
24 solar radii, while sophisticated numerical solutions
(Keppens \& Goedbloed 2000, their Fig.\ 3) yield 7 to 14 stellar radii in the
segment of their solution with open field lines -- where the stellar wind
comes from.  Recent observations for the 
Sun (Velli, Tenerani \& DeForest 2016) also indicate that for the Sun the
Alfv\'{e}n radius is of the order of 12 radii out over the polar regions. So
there is broad agreement between observations, theory and simulations
here.

However, it turns out that the expression of Eq.\ (3.1) as defining the stellar
wind torque is strongly dependent upon the geometry of the stellar coronal
magnetic field.  While the field in Weber \& Davis (1967) is purely radial 
(reversing at the equator), that in Keppens \& Goedbloed (2000) has a
much more realistic ``dead zone" over the equator, where the field is
closed, while open field lines spread out from the poles.  The dead
zone in the simulations takes on a form very similar to observed solar 
helmet streamers, as observed during a solar eclipse.

The torque from the stellar wind in Keppens \&
Goedbloed is a factor 15 to 60 smaller than that given by Eq. (3.1) (a
factor 10 to 40 compared to Weber \& Davis), because of the
difference in magnetic field topology.  The same result had already
been pointed out by Priest \& Pneuman (1974), based on the helmet
streamer geometry of Pneuman \& Kopp (1971).

I will show now that this result has important implications both for the 
mass loss required for the Sun to slow down from its initial rotation rate
at the ZAMS, to its current rate, and for the slow down of solar rotation
in the remainder of the Sun's main sequence lifetime.

First we need to know the moment of inertia of a star to be able to
estimate the slow down rate for a given stellar mass loss rate.  The
angular momentum of a star is given by
\begin{equation}
L_\star = I_{\star}\omega\, =\,  M_\star (\beta_I R_{\star})^2\omega,
\end{equation}
where $L_{\star}$ is the angular momentum, $I_{\star}$ the moment
of inertia, $\omega$ the rotation rate, $M_\star$ the stellar mass,
$R_{\star}$ its radius, and $\beta_I R_{\star}$ the radius of gyration, 
with $\beta_I$ the gyration
constant, i.e.\ the fraction of the radius for the arm in the moment of
inertia.  Stellar evolution codes show that the value of $\beta_I$ decreases
a little after arrival of a star at the ZAMS, because of the production
of Helium from lighter Hydrogen in the core (see Schrijver \& Zwaan 2000,
their Sect.\ 13.1).
Interestingly then we would expect Sun-like stars to slowly spin up as
they evolve on the main sequence if it weren't for stellar mass loss.  Later,
as the star evolves towards its giant phase, the moment of inertia greatly
increases of course, but that is of no concern here.  The typical value of
$\beta_I$ for a Sun-like star on the main sequence is of the order of 0.25,
the value I shall use from here on, and assumed to decrease much less 
than the rotation rate.

The decrease in angular momentum of a star as it evolves. i.e.\ 
$\dot{L_{\star}}$ is of course equal to the torque applied to it by the
stellar wind, i.e.\
\begin{equation}
\dot{L_{\star}} = M_\star (\beta_I R_{\star})^2\dot{\omega} = f  \omega
R_{A}^{2}  \dot{M},
\end{equation}
where $f$ is the efficiency factor discussed above, determined by Keppens
\& Goedbloed (2000) to range from 1/60 to 1/15.  When we write the Alfv\'{en}
radius $R_{A}$ as a multiple of the stellar radius, $\alpha_{A} R_{\star}$ a
very simple expression results for the stellar mass loss that is required to 
produce the much better observed spin-down of late type stars after arriving
on the main sequence,
\begin{equation}
\dot{M} = \frac{\beta_{I}^2}{f {\alpha_{A}}^2} \frac{\dot{\omega}}{\omega}
M_{\star}.
\end{equation}

The term $\frac{\dot{\omega}}{\omega}$ is simply the inverse of the e-folding
time for the slow down in rotation of late type stars, which is of the order of
2-3 billion years, with not much variation between different stars (e.g.\ Nandy \&
Martens, 2007).  Above I have found $\beta_{I} \approx 0.25$, $\alpha_{A}
\approx$ 10, and $f \approx 1/30$.  Inserting these values into Eq.\ (3.4) we
derive our main result,
\begin{equation}
\dot{M_{\star}} = -7.5 \times 10^{-12} M_{\star} yr^-1.
\end{equation}

This represents the mass loss required to explain the observed stellar spin
down rate  by magnetic breaking.  This mass loss rate also equals the
mass loss required to resolve the Faint Young Sun Paradox, as discussed
in the previous section.  Indeed a very large mass loss, sustained for
several billion years, is not just a possibility, but it may very well be required
to explain the spin down of Sun-like stars.  The analysis above also
demonstrates that at its current mass loss rate of 2-3$\times$10$^{-14}$
M$_{\odot}$ yr$^{-1}$ our Sun will not slow down significantly for the
remainder of its presence on the main sequence.  This appears consistent
with the observed rotation rates of older late type stars on the main
sequence (Egeland, these proceedings).

\section{Discussion and Conclusions}  

\begin{figure}[ht]
\begin{center}
\includegraphics[width=0.47\linewidth]{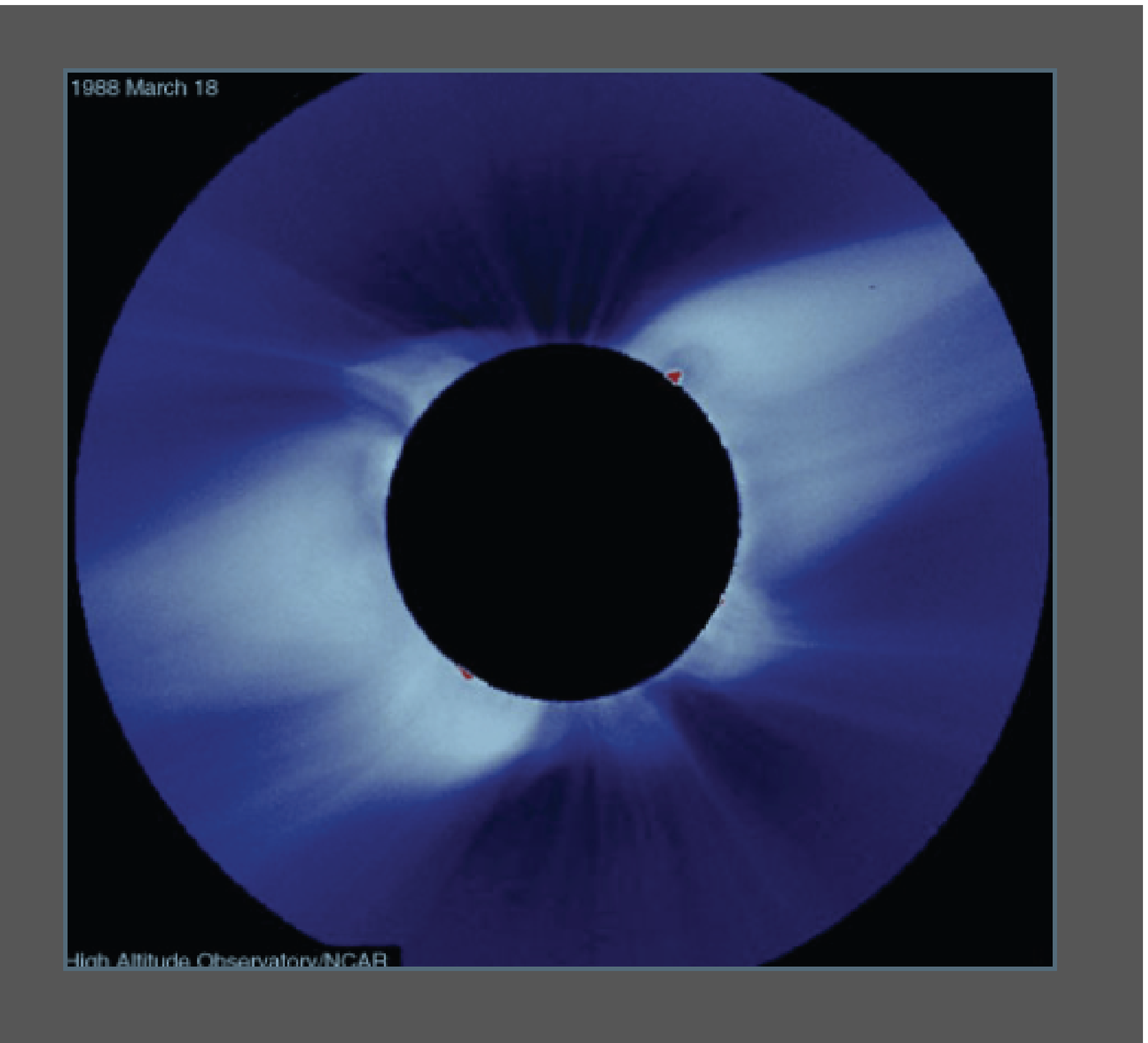}\hfill 
\includegraphics[width=0.48\linewidth]{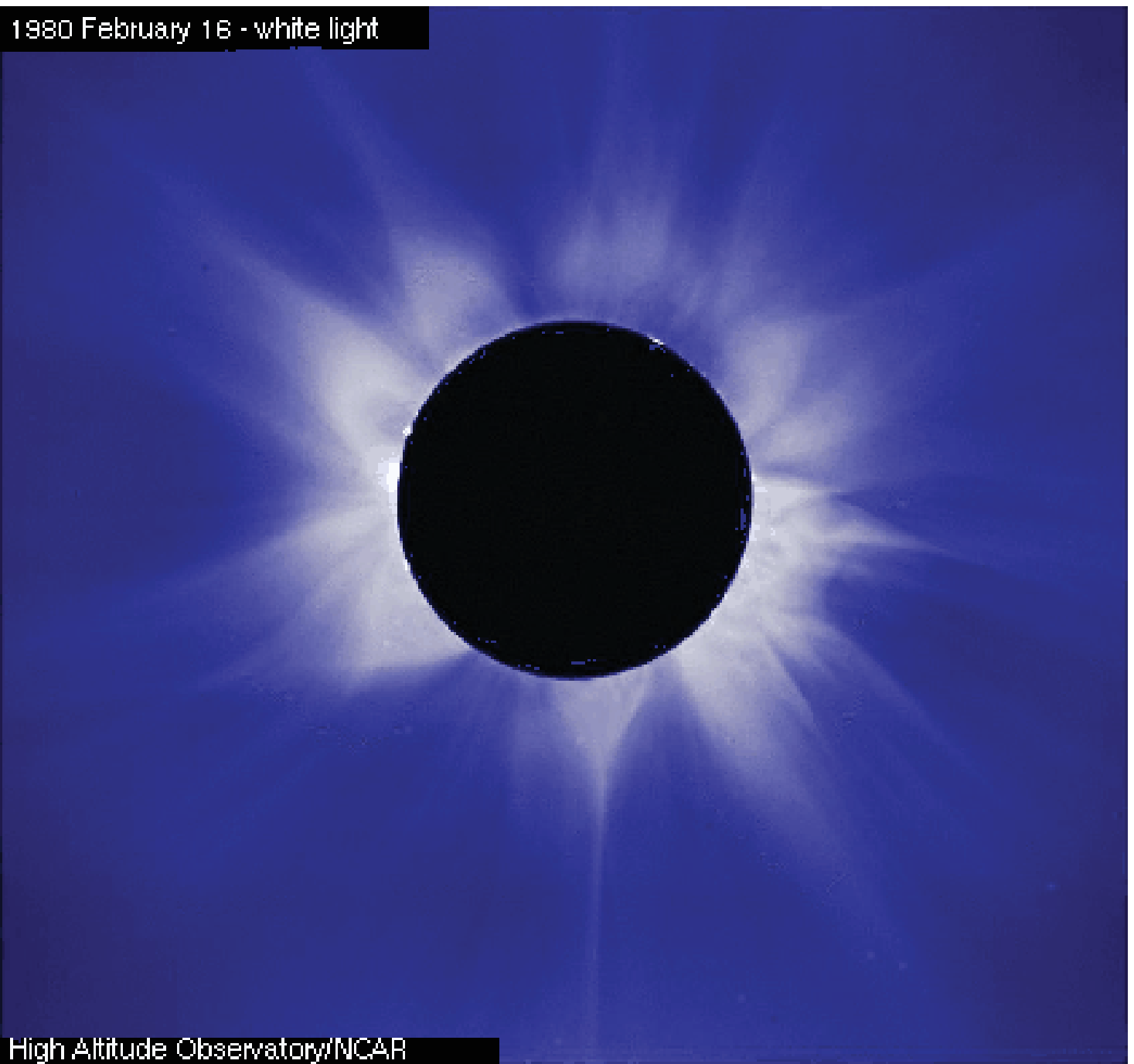}
 \caption{Helmet streamers observed during solar eclipses at solar minimum (left)
and solar maximum (right).  During maximum more streamers are present but their
size is smaller than the ones at minimum.}
   \label{fig1}
\end{center}
\end{figure}

The current mass loss rate of 2-3$\times$10$^{-14}$ M$_{\odot}$ yr$^{-1}$ is
not sufficient to slow down the Sun from an initial rotation period of 4-5 days
to its current value :  That would require an Alfv\'{en} radius of 170 solar
radii, beyond the orbit of Venus.   One might argue that the young Sun most
likely had a much stronger magnetic field, which would increase its Alfv\'{en}
radius.  However, the simulations of Keppens \& Goedbloed show that the
slow down is not very sensitive to magnetic field strength:  A stronger
magnetic field is compensated by a larger dead zone, keeping the wind 
torque nearly constant.   

Observations during solar eclipses even suggest a
shrinking Alfv\'{en} radius with solar magnetic field.  Fig.\ 1 shows juxtaposed
helmet streamers observed during eclipses near solar minimum and solar
maximum.  The solar maximum image on the right shows more helmets, as
expected, but of smaller size, with in particular the peak of the helmets closer
to the solar surface.  If, as in the simulations of Keppens \& Goedbloed, the
peak of the helmets approximately coincides with the Alfv\'{en} radius at that
position angle, the Alfv\'{en} radius of the Sun, in its current phase of evolution,
is indeed smaller at higher activity levels. 

I conclude that a high mass loss rate is a 
reasonable hypothesis to resolve both the Faint Young Sun and the Faint
Young Stars Paradoxes.  Observations of mass loss rates of late type
stars in the first billions of years after their arrival on the ZAMS are needed 
to verify this hypothesis, as well as, if possible, investigations of remaining
signatures of the early solar wind.


\begin{thebibliography}{}

\bibitem[Kasting(1989)]{1989GPC.....1...83K} Kasting, J.~F.\ 1989, {\it Global and Planetary Change}, 1, 83

\bibitem[Kasting(1991)]{1991Icar...94....1K} Kasting, J.~F.\ 1991, \icarus, 94, 1

\bibitem[Kasting(2004)]{2004AGUFM.B33E..01K} Kasting, J.~F.\ 2004, {\it AGU Fall Meeting Abstracts}

\bibitem[Kasting et al.(1988)]{1988SciAm.258e..46K} Kasting, J.~F., Toon, O.~B., \& Pollack, J.~B.\ 1988, {\it Scientific American}, 258, 46

\bibitem[Keppens \& Goedbloed(2000)]{2000ApJ...530.1036K} Keppens, R., \& Goedbloed, J.~P.\ 2000, \apj, 530, 1036

\bibitem[Minton \& Malhotra(2007)]{2007ApJ...660.1700M} Minton, D.~A., \& Malhotra, R.\ 2007, \apj, 660, 1700

\bibitem[Nandy \& Martens(2007)]{2007AdSpR..40..891N} Nandy, D., \& Martens, P.~C.~H.\ 2007, {\it Advances in Space Research}, 40, 891

\bibitem[do Nascimento et al.(2016)]{2016ApJ...820L..15D} do Nascimento, J.-D., Jr., Vidotto, A.~A., Petit, P., et al.\ 2016, \apjl, 820, L15

\bibitem[Pneuman \& Kopp(1971)]{1971SoPh...18..258P} Pneuman, G.~W., \& Kopp, R.~A.\ 1971, \solphys, 18, 258

\bibitem[Priest \& Pneuman(1974)]{1974SoPh...34..231P} Priest, E.~R., \& Pneuman, G.~W.\ 1974, \solphys, 34, 231

\bibitem[Sagan \& Mullen(1972)]{1972Sci...177...52S} Sagan, C., \& Mullen, G.\ 1972, {\it Science}, 177, 52 

\bibitem[Schrijver \& Zwaan(2000)]{2000ssma.book.....S} Schrijver, C.~J., \& Zwaan, C.\ 2000, Solar and stellar magnetic activity / Carolus J.~Schrijver, Cornelius Zwaan.~ New York : Cambridge University Press, 2000.~(Cambridge astrophysics series ; 34) 

\bibitem[Velli et al.(2016)]{2016SPD....4740205V} Velli, M., Tenerani, A., \& DeForest, C.\ 2016, AAS/{\it Solar Physics Division Meeting}, 47, 402.05 

\bibitem[Weber \& Davis(1967)]{1967ApJ...148..217W} Weber, E.~J., \& Davis, L., Jr.\ 1967, \apj, 148, 217 

\end{thebibliography}
\end{document}